# Prefrontal electrical stimulation in nondepressed reduces levels of reported negative affects from daily stressors

Adelaide Austin [a], Gabriela Jiga-Boy [a,c], Sara Rea [a], Simon Newstead [a], Sian Roderick [b,c], Nick J Davis [a], R. Marc Clement [b], and Frédéric Boy [a,b,c]

[a] Department of Psychology, College of Human and Health Science, Swansea University, Wales | [b] Scientia Research Group | School of Medicine, Swansea University, Wales | [c] NeuroTherapeutics Limited, Institute of Life Science, Swansea University, SA2 8PP Swansea, Wales,

# Abstract

Negative emotional responses to the daily life stresses have cumulative effects which, in turn, impose wide-ranging negative constraints on emotional well being and neurocognitive performance (Kalueff:2007cp, Charles:2013eq, Nadler:2010hk). Crucial cognitive functions such as memory and problem solving, as well more short term emotional responses (e.g., anticipation of- and response to- monetary rewards or losses) are influenced by mood. The negative impact of these behavioural responses is felt at the individual level, but it also imposes major economic burden on modern healthcare systems. Although much research have been undertaken to understand the underlying mechanisms of depressed mood and design efficient treatment pathways, comparatively little was done to characterize mood modulations that remain within the boundaries of a healthy mental functioning. In one placebo-controlled experiments, we applied daily prefrontal transcranial Direct Current Stimulation (tDCS) at five points in time, and found reliable improvements on self-reported mood evaluation. We replicated this finding in an independent double-blinded placebo-controlled experiment and showed that stimulation over a shorter period of time (3 days) is sufficient to create detectable mood improvements. Taken together, our data show that repeated bilateral prefrontal tDCS can reduce psychological distress in nondepressed individuals.

## 1. Introduction

One function of the dorsolateral prefrontal cortex (dlPFC) is to continuously appraise the emotional content of daily-life situations, and to rapidly regulate oriented responses (Levesque *et al.*, 2003, Banks *et al.*, 2007). The strong negative impact of daily stressors on current mood is well known (Bolger *et al.*, 1989). Over time, the outcomes of this idiosyncratic evaluative and responsive process amass, and impact individuals' emotional wellbeing and neurocognitive performance (Charles *et al.*, 2013, Nadler *et al.*, 2010). Here, we exploited the modulation of GABA- and glutamate-ergic neurotransmission (Stagg *et al.*, 2011a; Stagg *et al.*, 2011b; Stagg *et al.*, 2009, Kim *et al.*, 2014} and cortical excitability (Romero Lauro *et al.*, 2014) caused by tDCS to determine whether negative emotional responses to daily-life stresses can be reduced in healthy individuals. tDCS involves placing two macro-electrodes on the scalp, and passing a weak regulated direct current (in the order of the mA) between them. Recent evidence from clinical research shows that repeated prefrontal tDCS in depressed patients produces measurable clinical benefits. Meta-analyses of recent open-label studies and double-blinded trials for the treatment of major depressive disorder (Brunoni *et al.*, 2011a, DellOsso *et al.*, 2012, Boggio *et al.*, 2008; Fregni *et al.*, 2006a; Fregni *et al.*, 2006b, Loo *et al.*, 2012; Loo *et al.*, 2010), found that active prefrontal tDCS was associated, on average, with a 29.1% ± 4.6% reduction in depressive symptoms; and five of these studies detected long-lasting benefits a month after the last stimulation. In addition, Brunoni *et al.* (2011b) also found that 1- active tDCS was more effective than sham, 2- tDCS was as effective as Sertraline, a selective serotonin reuptake inhibitor (SSRI) antidepressant and, 3- tDCS and SSRI combined have greater efficacy than each treatment alone. This body of evidence strongly suggest that repeated daily prefrontal tDCS can be an effective tool for improving mood in depressed patient.

However, the present challenge is to understand the neurobiological underpinnings and the psychological mechanisms at play in this effect. Here, we took the original approach of studying how repeated prefrontal tDCS modulated the way nondepressed volunteers self-evaluated the emotional states consequent to life events (stressful or not). This is particularly relevant since one of the leading causal factors in depression onset is the accumulation of negative emotional states resulting from sustained or chronic exposure to stressful life events (Gandiga *et al.*, 2006).

## 2. Methods

a. **Participants.**

Sixty-six early-adults, unmedicated, nondepressed females from Swansea University (mean age: 21.6 years ± 2.3, mainly Caucasian) participated in the experiments reported here in exchange of payment (£20) or course credit. All were naïve to the purpose of the experiments and had no neuropsychiatric history. Participants were aware that the experimental manipulation repeatedly used tDCS neuromodulation and that they would have to complete several questionnaires, but no further specification was given as to the nature of the hypotheses. The departmental Research Ethics Committee approved all procedures. After completion of experiment 1, two participants voluntarily reported significant events that affected their current mood (passing of a relative, relationship breakup), and their data were discarded.

b. **Bilateral prefrontal tDCS.**

A DC stimulator (HDCStim, Newronika, Milano, Italy) delivered a 1500 mA current to the scalp via 5x5 cm rubber-graphite electrodes (current density: 0.06 mA/cm$^2$). Impedance was automatically monitored every 5s, and tension adjusted accordingly, so that to deliver constant current (within safety limits). In experiment 1 and 2 the anode was centred over the left F3 10-20 position (see Fig. 1). The cathode was placed over the contralateral F4 position (for similar electrodes placement see (Brunoni *et al.*, 2011b, Dell'Osso *et al.*, 2012). Sponges soaked in 0.9% NaCl solution (Sterowash, Steroplast, Manchester, UK) were used to create a conducting medium between the scalp and electrodes. For active stimulation, the current was ramped up over 15 seconds and was then held at 1500mA for 12 minutes, before being ramped down over 15s. For sham stimulations, the stimulator was automatically switched off after an initial ramp-up (15s at 0.1 mA.s$^{-1}$), plateau periods (6s at 1.5 mA), and final ramp-down (15s at -0.1 mA.s$^{-1}$), to create a realistic placebo control condition (Brunoni *et al.*, 2012) that still generate the short lasting tingling sensations identical to that felt at the beginning of the active tDCS stimulations. Usually, only these very mild sensations are experienced (Gandiga *et al.*, 2006), and when directly asked, most participants do not even perceive a difference between active and sham stimulation (Poreisz *et al.*, 2007}. Experiment 1 was single blinded, whereas Experiment 2 was a double-blind randomized trial, where neither participants, nor experimenters knew whether the stimulation was active or sham.

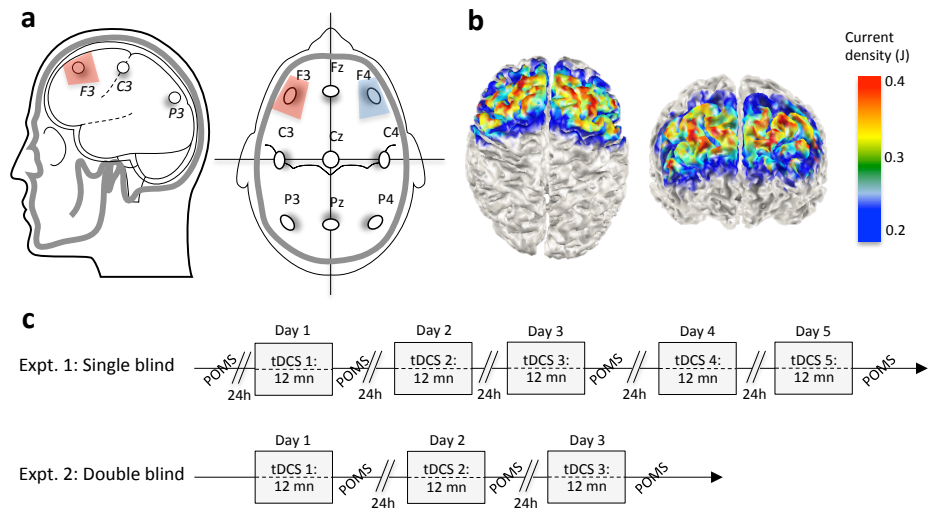

Figure 1 | a, Positioning of the stimulating scalp electrodes according to the 10-20 system nomenclature, and in reference to the main cortical fissures b, 3D numerical computation of electric fields on the surface of a cortical model for 5 x 5 cm electrodes placed on F3 and F4 head locations (Jung *et al.*, 2013) c, Timelines of experiments.

### c. Mood Assessment

The Profile Of Mood States (Pollock *et al.*, 1979) questionnaire provides a rapid method of assessing transient, fluctuating active mood states. It is an instrument that is particularly well suited to the present research because of its sensitivity to change in affective states. We used the abridged scale - a 24-item questionnaire that measures mood along six dimensions: tension-anxiety, depression-dejection, anger-hostility, vigour-activity, fatigue-inertia, and confusion-bewilderment (Curran et al, 2006). Participants rated how they were currently feeling with respect to 24 words (e.g., Worn-out, Annoyed, Confused, Active, Panicky, Unhappy) on a scale of 1 ("Not at all") to 5 ("Extremely"). Scores at each of the factor scores, except for the vigour-activity score, was added together; and then, the vigour-activity score was then subtracted from this total to produce a general composite mood score. In Experiment 1, although participants received tDCS daily over five days, we limited the number of post-stimulation mood assessments by only administering the POMS every other day. In Experiment 2, where participants were stimulated daily over three days, we administered the POMS immediately after every stimulation (see Fig 1).

## 3. Results

In two experiments, we present converging evidence that series of daily bilateral prefrontal tDCS sessions positively impacted the self-assessment of mood states. In Experiment 1, we first established that, when five 12 mn daily tDCS sessions were administered, scores at the Profile of Mood States scale were improved in the active ($p < 1.1e-05$, Fig. 1a), but not in the sham condition ($p = NS$, Fig. 1b). In the active condition, significant improvements were found between evaluations carried out each other day (all $p$s $< 0.01$), whereas no change was noted between sham sessions ($p = NS$). This striking dichotomy was independently replicated in Experiment. 2, where tDCS sessions were administered on three consecutive days (active: $p < 1.55e-06$, Fig. 2c; sham: $p = NS$, Fig. 2d). In substantive terms, the reduction in negative mood states, in the two active tDCS conditions, accounted for 64.7% and 39.1% of the total variations in scores in Expt. 1 and 2, respectively.

The absence of significant mood changes in the sham condition, where participants received series of 36s 1.5 mA daily stimulations, insured that the observed negative mood reduction was not due to a learning or habituation effect, with participants (consciously or unconsciously) gradually providing less negative ratings during the mood evaluation.

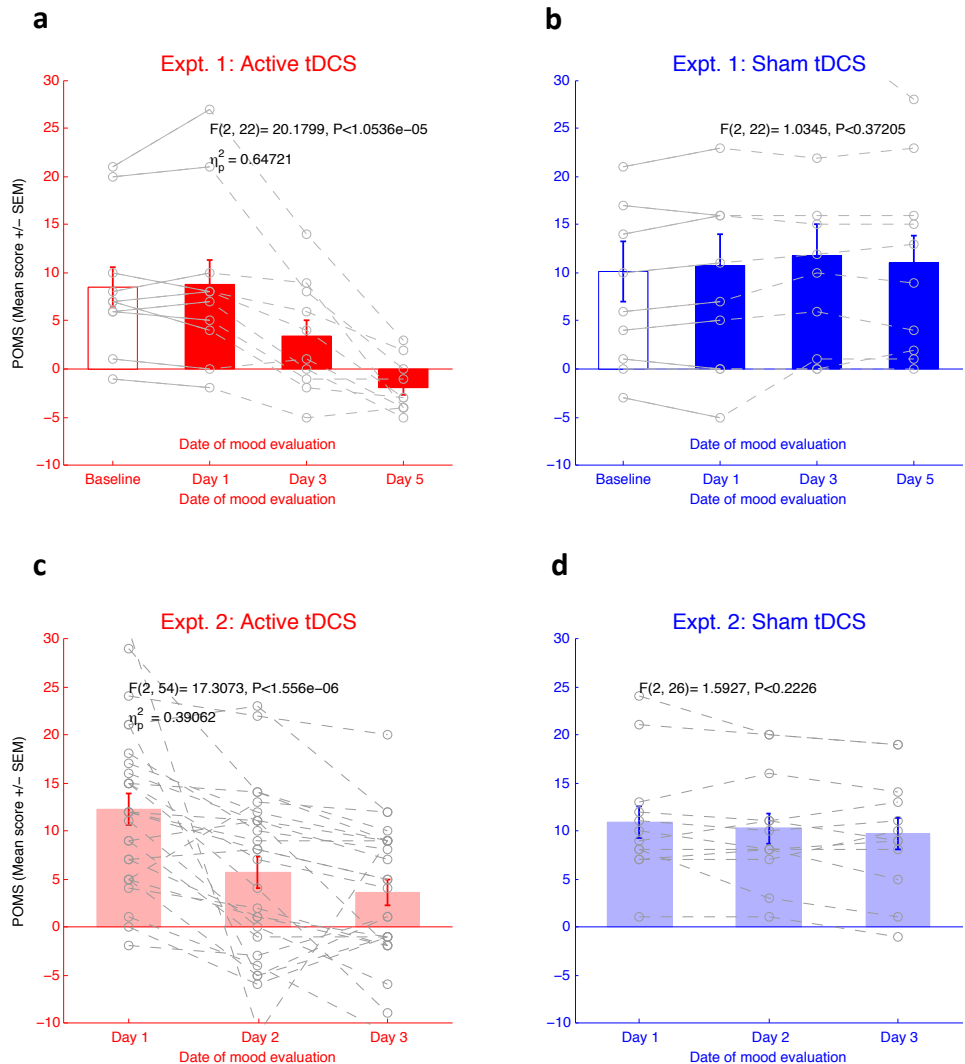

Figure 2 | a, b, Evolution of mood states self-evaluation (total score) throughout the three days of brain stimulation in the active, and the "sham" conditions. Grey line present individual performances in each condition. c, d, Similar plots for replication experiment 2.

The general tendency towards mood improvement during active tDCS evidenced in the reduction in general composite mood score is logically resulting from improvements in each of the subscales. Although the design of the present research is not adapted to such subsampling of the data, we decided to still present how scores at each of the six subscales in the POMS were modulated by tDCS, without presenting any result of statistical testing (Fig. 3). Although the argument is only descriptive, and variability is high, we note that, for each scale except "vigour", there is an amelioration tendency (a decrease in scores) in the active tDCS but not in the Sham condition. Our current research programme explores this aspect of these datasets, in an adapted research protocol with sufficient statistical power.

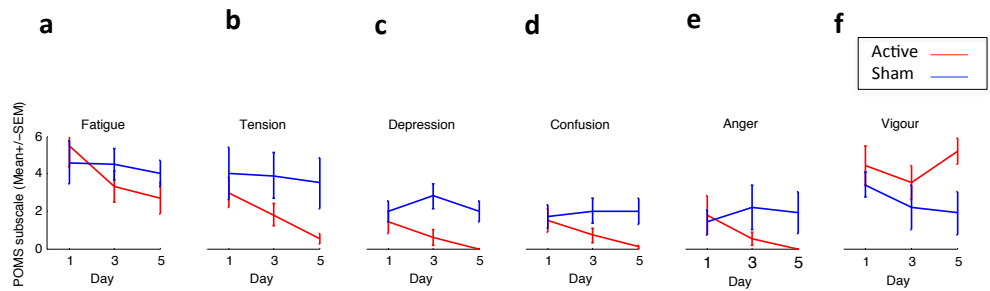

Figure 3 | a:f, Evolution of self-evaluation for each dimensions of the POMS throughout the three days of brain stimulation in the active and sham conditions in experiment 1.

## 4. Discussion

In two sham-controlled experiments, we found that repeated daily prefrontal tDCS sessions over 5 several days could effectively modulate how nondepressed individuals self assess their mood states. Results show that participants experienced less psychological distress from daily stressors, a well established cause in the establishment of a negative emotional state (Bolger *et al.*, 1989). We replicated this finding in an independent, randomized, double-blind experiment applying similar protocol and stimulation on 3 consecutive days.

To our knowledge, the present research is the first to show that the amount negative mood states, in unmedicated nondepressed individuals, can be reduced with repeated prefrontal tDCS. This is consistent with prior clinical body of research demonstrating that repeated tDCS improves depressive symptoms. Although, there are limited by a small number of studies with small sample size, the most recent systematic reviews to date (Meron *et al.*, 2015; Kalu *et al.*, 2012) reach nonetheless the conclusion that tDCS significantly reduces symptoms of major depressive episodes. It is also true that a few conflicting studies failed to find a reduction in psychological distress following prefrontal tDCS. However, the body of research in question either examined individuals with treatment-resistant depression, or else participant samples that were concurrently taking various medication treatments known to interact with tDCS (Loo *et al.*, 2010; Bennabi *et al.*, 2015, Blumberger *et al.*, 2012). For example, the administration of GABA-agonist benzodiazepines (Lorazepam) delays, enhances and prolongs the elevation in cortical excitability resulting from anodal tDCS (Nitsche *et al.*, 2004), while Serotonin Selective Reuptake Inhibitors such as Citalopram concurrently increases anodal effects, and transforms cathodal inhibition into facilitation (Bennabi *et al.*, 2015, Nitsche *et al.* 2009, Brunoni *et al.*, 2013).

Only a few other studies have examined the possibility of modulating mood using tDCS in healthy individuals, these however failed to detect notable effect (Koenigs *et al.*, 2009, Plazier *et al.*, 2012, Morgan *et al.*, 2014). While it is difficult to discuss the absence of notable effect, we believe these could be accounted for by radically different research designs, stimulation programme, electrode montages, or in the way current mood was evaluated. For example, in both experiments reported here, participants underwent either 5 or 3 consecutive days of active (or sham) bifrontal tDCS, and our conclusions are therefore founded upon comparisons between an active group and a sham group, and self reported modulation of mood occurring across days. In contrast, Plazier *et al.*'s (2012) goals were rather different and the research was looking for alterations in mood, following a single session of tDCS, utilising six forms of bifrontal and bioccipital stimulation, upon the same participant. Although such an attempt is both interesting and commendable, it is difficult to conceive that biochemical alterations within the cortex or detectable effects on mood resulting from a single 20mn tDCS session would be of similar origins to mood modulation observed across days of repeated stimulation. Of importance, we also think that Plazier *et al.*'s (2012) way of administering mood questionnaires directly before and after the stimulation, is far from optimal: The short time period between repeated assessments, and a participants' initial responses will likely have influenced, to some degree, latter responses to the questionnaires re-administered directly following stimulation. This observation regarding the limits of repeated questionnaire administration also applies to Morgan *et al.* (2014) and Koenigs *et al.*, (2009) for the Positive Affect Negative Affect Schedule (PANAS), and the POMS, respectively. Of importance, Koenigs *et al.*, (2009) rather placed the anode on the frontal poles bilaterally ($F_{p1}$ and $F_{p2}$), and used an extracephalic reference electrode, but detected no mood improvement after any of three usual tDCS conditions

Discussion of certain methodological questions is needed to further inform future investigations. For instance, our data may have implications for the interpretation of numerous findings in which prefrontal tDCS induces cognitive improvements (e.g., Kadosh *et al.*, 2010, Jacobson *et al.*, 2011}. Cognitive processing is affected by mood, with positive mood being associated with improved cognitive performance (Nadler *et al.*, 2010), and since we show that tDCS reduces self reported psychological distress, it is possible that tDCS-induced cognitive improvement are actually mediated by a mood improvement (or vice-versa). Current neuromodulatory work in our team address this issue, and aims at disentangling the complex interaction between mood and cognitive performance. Another pertinent issue relates to the duration of the tDCS-induced mood modulation. In clinical studies, researchers have reported mood improvement effects to be maintained for at least one month after the last stimulation (Kalu *et al.*, 2012. However, these studies involve a greater number of stimulation sessions (N = 10), over a longer period of time (two weeks), and that indicates that the optimal programme of stimulation needed to warrant a potentiated reduction of psychological distress in a non-clinical population still has to be determined.

Both past studies in depression, and the present work indicate, respectively, that tDCS is effective in reducing depressive symptoms and psychological distress, respectively. An important question to consider concerns the identification of the neurophysiological mechanisms that are able to induce these changes. One possible explanation follows from two programmes of research. One that examined the relationship between GABA levels and depression, and evidenced that GABA-agonist drugs and agents all tend to ameliorate the depressive symptomatology in human, and in animals (Kalueff *et al.*, 2007). Another, more recent body of research, showed that tDCS could lower cortical GABA and Glutamate level locally (Stagg *et al.*, 2009). Although the latter effect was obtained in regions of the frontal cortex that are not directly causally related to mood regulation, unlike the dlPFC, we believe that it provides a general framework for the generation of testable hypotheses. The details of the interaction between GABAergic neuromodulation within the dlPFC and mood regulation are likely to be complex, as the dlPFC forms part of a network involving loops through striatum and thalamus as well as numerous connections to other cortical and subcortical areas relevant for regulating mood. Similarly, the apparent contradiction between the effect of a technique which lowers levels of GABA and a general GABA deficit theory in depression has to be understood in this context, and treated with great care. It is indeed possible that the tDCS-induced local reduction in GABA concentration results in potentiated GABAergic neurotransmission along these extended networks (for similar reasoning, see discussion in Boy *et al.*, 2011).

**Conflict of Interest Statement**

FB filed the patent application no. 1503004.2 (Intellectual Property Office, Newport, UK) for a novel tDCS device. GJB, SR and FB hold shares in NeuroTherapeutics Limited, a UK registered Company. RMC is an indirect shareholder. The other authors declare that the research was conducted in the absence of any commercial or financial relationships that could be construed as a potential conflict of interest. At the time of data collection, students who gave tDCS stimulation and collected behavioural responses were blind to the hypotheses and previous findings.

**Aknowledgement**

The authors wish to thank the BIAL Foundation for funding this research, and value the efforts of the students who collected the data (in particular Cathy Ghalib and Martin Stevenson)

**Reference**

Banks, S. J., Eddy, K. T., Angstadt, M., Nathan, P. J., & Phan, K. L. (2007). Amygdala-frontal connectivity during emotion regulation. *Social Cognitive and Affective Neuroscience*, *2*(4), 303–312. doi:10.1093/scan/nsm029

Bennabi, D., Nicolier, M., Monnin, J., Tio, G., Pazart, L., Vandel, P., & Haffen, E. (2015). Pilot study of feasibility of the effect of treatment with tDCS in patients suffering from treatment-resistant depression treated with escitalopram. *Clinical Neurophysiology : Official Journal of the International Federation of Clinical Neurophysiology*, *126*(6), 1185–1189. doi:10.1016/j.clinph.2014.09.026

Blumberger, D. M., Tran, L. C., Fitzgerald, P. B., Hoy, K. E., & Daskalakis, Z. J. (2012). A randomized double-blind sham-controlled study of transcranial direct current stimulation for treatment-resistant major depression. *Frontiers in Psychiatry*, *3*, 74. doi:10.3389/fpsyt.2012.00074

Boggio, P. S., Rigonatti, S. P., Ribeiro, R. B., Myczkowski, M. L., Nitsche, M. A., Pascual-Leone, A., & Fregni, F. (2008). A randomized, double-blind clinical trial on the efficacy of cortical direct current stimulation for the treatment of major depression. *The International Journal of Neuropsychopharmacology / Official Scientific Journal of the Collegium Internationale Neuropsychopharmacologicum (CINP)*, *11*(2), 249–254. doi:10.1017/S1461145707007833

Bolger, N., DeLongis, A., Kessler, R. C., & Schilling, E. A. (1989). Effects of daily stress on negative mood. *Journal of Personality and Social Psychology*, *57*(5), 808–818.

Boy, F., Evans, C. J., Edden, R. A. E., Lawrence, A. D., Singh, K. D., Husain, M., & Sumner, P. (2011). Dorsolateral Prefrontal γ-Aminobutyric Acid in Men Predicts Individual Differences in Rash Impulsivity. *Biological Psychiatry*, *70*(9), 866–872. doi:10.1016/j.biopsych.2011.05.030

Brunoni, A. R., Valiengo, L., Baccaro, A., Zanao, T. A., de Oliveira, J. F., Vieira, G. P., *et al.* (2011a). Sertraline vs. Electrical Current Therapy for Treating Depression Clinical Trial - SELECT TDCS: Design, rationale and objectives. *Contemporary Clinical Trials*, *32*(1), 90–98. doi:10.1016/j.cct.2010.09.007

Brunoni, A. R., Ferrucci, R., Bortolomasi, M., Vergari, M., Tadini, L., Boggio, P. S., *et al.* (2011b). Transcranial direct current stimulation (tDCS) in unipolar vs. bipolar depressive disorder. *Progress in Neuro-Psychopharmacology & Biological Psychiatry*, *35*(1), 96–101. doi:10.1016/j.pnpbp.2010.09.010

Brunoni, A. R., Nitsche, M. A., Bolognini, N., Bikson, M., Wagner, T., Merabet, L., *et al.* (2012). Clinical research with transcranial direct current stimulation (tDCS): Challenges and future directions. *Brain Stimulation*, *5*(3), 175–195. doi:10.1016/j.brs.2011.03.002

Brunoni, A. R., Valiengo, L., Baccaro, A., Zanão, T. A., de Oliveira, J. F., Goulart, A., *et al.* (2013). The Sertraline vs Electrical Current Therapy for Treating Depression Clinical Study. *JAMA Psychiatry*, *70*(4), 383. doi:10.1001/2013.jamapsychiatry.32

Charles, S. T., Piazza, J. R., Mogle, J., Sliwinski, M. J., & Almeida, D. M. (2013). The Wear and Tear of Daily Stressors on Mental Health. *Psychological Science*, *24*(5), 733–741. doi:10.1177/0956797612462222

Curran, S. L., Andrykowski, M. A., & Studts, J. L. (2004). Short Form of the Profile of Mood States (POMS-SF): Psychometric information. *Psychological Assessment*, *7*(1), 1–4.

Dell'Osso, B., Zanoni, S., Ferrucci, R., Vergari, M., Castellano, F., D'Urso, N., *et al.* (2012). Transcranial direct current stimulation for the outpatient treatment of poor-responder depressed patients. *European Psychiatry*, *27*(7), 513–517. doi:10.1016/j.eurpsy.2011.02.008

Fregni, F., Boggio, P. S., Nitsche, M. A., Marcolin, M., Rigonatti, S. P., & Pascual-Leone, A. (2006a). Treatment of major depression with transcranial direct current stimulation. *Bipolar Disorders*, *8*, 203–204.

Fregni, F., Boggio, P. S., Nitsche, M. A., Rigonatti, S. P., & Pascual-Leone, A. (2006b). Cognitive effects of repeated sessions of transcranial direct current stimulation in patients with depression. *Depression and Anxiety*, *23*(8), 482–484. doi:10.1002/da.20201

Gandiga, P. C., Hummel, F. C., & Cohen, L. G. (2006). Transcranial DC stimulation (tDCS): a tool for double-blind sham-controlled clinical studies in brain stimulation. *Clinical Neurophysiology : Official Journal of the International Federation of Clinical Neurophysiology*, *117*(4), 845–850. doi:10.1016/j.clinph.2005.12.003

Jacobson, L., Javitt, D. C., & Lavidor, M. (2011). Activation of inhibition: diminishing impulsive behavior by direct current stimulation over the inferior frontal gyrus. *Journal of Cognitive Neuroscience*, *23*(11), 3380–3387. doi:10.1162/jocn_a_00020

Kadosh, R. C., Soskic, S., Iuculano, T., & Kanai, R. (2010). Modulating neuronal activity produces specific and long-lasting changes in numerical competence. *Current Biology 20, 2016–2020.*

Kalu, U. G., Sexton, C. E., Loo, C. K., & Ebmeier, K. P. (2012). Transcranial direct current stimulation in the treatment of major depression: a meta-analysis. *Psychological Medicine*, *42*(09), 1791–1800. doi:10.1017/S0033291711003059

Kalueff, A. V., & Nutt, D. J. (2007). Role of GABA in anxiety and depression. *Depression and Anxiety*, *24*(7), 495–517. doi:10.1002/da.20262

Kim, S., Stephenson, M. C., Morris, P. G., & Jackson, S. R. (2014). tDCS-induced alterations in GABA concentration within primary motor cortex predict motor learning and motor memory: A 7T magnetic resonance spectroscopy study. *Neuroimage*, *99*(C), 237–243. doi:10.1016/j.neuroimage.2014.05.070

Koenigs, M., Ukueberuwa, D., Campion, P., Grafman, J., & Wassermann, E. (2009). Bilateral frontal transcranial direct current stimulation: Failure to replicate classic findings in healthy subjects. *Clinical Neurophysiology : Official Journal of the International Federation of Clinical Neurophysiology*, *120*(1), 80–84. doi:10.1016/j.clinph.2008.10.010

Lévesque, J., Eugène, F., Joanette, Y., Paquette, V., Mensour, B., Beaudoin, G., *et al.* (2003). Neural circuitry underlying voluntary suppression of sadness. *Biological Psychiatry*, *53*(6), 502–510.

Loo, C. K., Sachdev, P., Martin, D., Pigot, M., Alonzo, A., Malhi, G. S., *et al.* (2010). A double-blind, sham-controlled trial of transcranial direct current stimulation for the treatment of depression. *International Journal of Neuropsychopharmacology*, *13*(1), 61–69. doi:10.1017/S1461145709990411

Loo, C. K., Alonzo, A., Martin, D., Mitchell, P. B., Galvez, V., & Sachdev, P. (2012). Transcranial direct current stimulation for depression: 3-week, randomised, sham-controlled trial. *The British Journal of Psychiatry*, *200*(1), 52–59. doi:10.1192/bjp.bp.111.097634

Meron, D., Hedger, N., Garner, M., & Baldwin, D. S. (2015). Transcranial direct current stimulation (tDCS) in the treatment of depression: Systematic review and meta-analysis of efficacy and tolerability. *Neuroscience and Biobehavioral Reviews*, *57*, 46–62. doi:10.1016/j.neubiorev.2015.07.012

Morgan, H. M., Davis, N. J., & Bracewell, R. M. (2014). Does transcranial direct current stimulation to prefrontal cortex affect mood and emotional memory retrieval in healthy individuals. *PLoS ONE, 9(3), e92162.*

Nadler, R. T., Rabi, R., & Minda, J. P. (2010). Better Mood and Better Performance: Learning Rule-Described Categories Is Enhanced by Positive Mood. *Psychological Science*, *21*(12), 1770–1776. doi:10.1177/0956797610387441

Nitsche, M. A., Liebetanz, D., Schlitterlau, A., Henschke, U., Fricke, K., Frommann, K., *et al.* (2004). GABAergic modulation of DC stimulation-induced motor cortex excitability shifts in humans. *The European Journal of Neuroscience*, *19*(10), 2720–2726. doi:10.1111/j.0953-816X.2004.03398.x

Nitsche, M. A., Kuo, M.-F., Karrasch, R., Wächter, B., Liebetanz, D., & Paulus, W. (2009). Serotonin Affects Transcranial Direct Current–Induced Neuroplasticity in Humans. *Biological Psychiatry*, *66*(5), 503–508. doi:10.1016/j.biopsych.2009.03.022

Poreisz, C., Boros, K., Antal, A., & Paulus, W. (2007). Safety aspects of transcranial direct current stimulation concerning healthy subjects and patients. *Brain Research Bulletin*, *72*(4-6), 208–214. doi:10.1016/j.brainresbull.2007.01.004

Plazier, M., Joos, K., Vanneste, S., Ost, J., & De Ridder, D. (2012). Bifrontal and bioccipital transcranial direct current stimulation (tDCS) does not induce mood changes in healthy volunteers: A placebo controlled study. *Brain Stimulation*, *5*(4), 454–461. doi:10.1016/j.brs.2011.07.005

Pollock, V., Cho, D. W., Reker, D., & Volavka, J. (1979). Profile of Mood States: the factors and their physiological correlates. *Journal of Nervous and Mental Disease*, *167*(10), 612–614.

Romero Lauro, L. J., Rosanova, M., Mattavelli, G., Convento, S., Pisoni, A., Opitz, A., *et al.* (2014). TDCS increases cortical excitability: direct evidence from TMS–EEG. *Cortex; a*

*Journal Devoted to the Study of the Nervous System and Behavior*, *58*, 99–111. doi:10.1016/j.cortex.2014.05.003

Stagg, C. J., Best, J. G., Stephenson, M. C., O'Shea, J., Wylezinska, M., Kincses, Z. T., *et al.* (2009). Polarity-Sensitive Modulation of Cortical Neurotransmitters by Transcranial Stimulation. *The Journal of Neuroscience*, *29*(16), 5202–5206. doi:10.1523/JNEUROSCI.4432-08.2009

Stagg, C. J., Bachtiar, V., & Johansen-Berg, H. (2011a). The Role of GABA in Human Motor Learning. *Current Biology*, *21*(6), 480–484. doi:10.1016/j.cub.2011.01.069

Stagg, C. J., & Nitsche, M. A. (2011b). Physiological Basis of Transcranial Direct Current Stimulation. *The Neuroscientist*, *17*(1), 37–53. doi:10.1177/1073858410386614